\documentclass[aps,pra, twocolumn, superscriptaddress]{revtex4-1}

\usepackage[margin=1in]{geometry}
\usepackage[T1]{fontenc}
\usepackage{bbm}
\usepackage{hyperref}
\hypersetup{%
	breaklinks=true,
	colorlinks,
	linkcolor={cyan},
	citecolor={cyan},
	urlcolor={cyan},
	pdfauthor = {Aniruddha Bapat, Stephen P. Jordan},
	pdftitle ={Approximate optimization of MAXCUT with a local spin algorithm},
	pdfpagemode=UseOutlines,
	pdfborder={0 0 0}
}

\usepackage{titlesec}
\titleformat{\section}[block]{\bfseries}{\thesection}{5pt}{\filcenter\MakeUppercase}{}
\titlespacing*{\section}{0pt}{*4}{*2}
\titleformat{\subsection}[runin]{\bfseries}{\thesubsection}{5pt}{\scshape}{}
\titlespacing*{\subsection}{0pt}{*1}{*2}

\usepackage{graphicx}
\graphicspath{{img/}}

\usepackage{amssymb,amsmath,amsthm}

\usepackage[capitalise]{cleveref}
\usepackage{booktabs, multirow, float}
\newcommand{\head}[1]{\textnormal{\textbf{#1}}}
\usepackage[toc,page, title,titletoc]{appendix}
\newcommand{\cutoff}[0]{\mathrm{cutoff}}
\newcommand{\paren}[1]{\left(#1\right)}
\newcommand{\brac}[1]{\left[#1\right]}
\newcommand{\curly}[1]{\left\{#1\right\}}
\newcommand{\mc}[1]{\mathcal{#1}}
\newcommand{\bra}[1]{\langle #1|}
\newcommand{\ket}[1]{|#1\rangle}
\newcommand{\braket}[2]{\langle #1|#2 \rangle}

\newcommand{\lrang}[1]{\langle #1\rangle}
\newcommand{\suml}[3]{\sum\limits_{#1}^{#2}#3}
\newcommand{\prodl}[3]{\prod\limits_{#1}^{#2}#3}

\newcommand{\id}[0]{\ensuremath{\mathbbm{1}}}

\renewcommand{\vec}[1]{\boldsymbol{\mathbf{#1}}}
\newcommand{\maxc}[0]{\textsc{maxcut}}
\newcommand{\lt}[0]{\textsc{lt}}

\begin{document}

\title{Approximate optimization of MAXCUT with a local spin algorithm}

\author{Aniruddha Bapat}
\email{ani@umd.edu}
\affiliation{Joint Center for Quantum Information and Computer Science, University of Maryland, College Park, Maryland 20742, USA}
\author{Stephen P. Jordan}
\email{stephen.jordan@microsoft.com}
\affiliation{Microsoft, Redmond, WA 98052, USA}
\affiliation{University of Maryland, College Park, MD 20742, USA}

\date{\today}

\begin{abstract}
Local tensor methods are a class of optimization algorithms that was introduced in [Hastings,  arXiv:1905.07047v2]\cite{hastings2019} as a classical analogue of the quantum approximate optimization algorithm (QAOA). 
These algorithms treat the cost function as a Hamiltonian on spin degrees of freedom and simulate the relaxation of the system to a low energy configuration using local update rules on the spins. 
Whereas the emphasis in \cite{hastings2019} was on theoretical worst-case analysis, we here investigate performance in practice through benchmarking experiments on instances of the \maxc{} problem. 
Through heuristic arguments we propose formulas for choosing the hyperparameters of the algorithm which are found to be in good agreement with the optimal choices determined from experiment. 
We observe that the local tensor method is closely related to gradient descent on a relaxation of \maxc{} to continuous variables, but consistently outperforms gradient descent in all instances tested. 
We find time to solution achieved by the local tensor method is highly uncorrelated with that achieved by a widely used commercial optimization package; on some \maxc{} instances the local tensor method beats the commercial solver in time to solution by up to two orders of magnitude and vice-versa. Finally, we argue that the local tensor method closely follows discretized, imaginary-time dynamics of the system under the problem Hamiltonian.
\end{abstract}

\maketitle
Binary unconstrained optimization, i.e., the maximization of an objective function on the configuration space of binary variables, is an important NP-hard optimization problem whose restrictions include several problems from Karp's list of 21 NP-complete problems~\cite{karp1972}. 
Due to the hardness of the problems, many solution approaches rely on finding approximately optima in the shortest possible time, or constructing algorithms that have an optimality guarantee but without guarantees on runtime. 
Algorithms in the latter category are often referred to as exact solvers, and include approaches that use linear, quadratic, or semi-definite programming (LP/QP/SDP) relaxations of the problem instance with techniques to obtain optimality bounds such as cutting planes, branch-and-bound, or Langrangian dual-based techniques. 
While the design of exact algorithms may be well-suited to theoretical analysis, the runtime scaling in instance size is often poor. 

In some cases, it is possible to design polynomial-time algorithms with guarantees on the \emph{approximation ratio}, i.e., the ratio of the optimum obtained to the global maximum. 
These are, however, ultimately limited by hardness results on achieving approximation ratios above a certain threshold value~\cite{haastad2001}. 

In the absence of runtime or optimality guarantees, problem-specific heuristics can nevertheless perform better than expected, exhibiting superior performance in runtime, optimality or both. 
An important class of heuristics takes inspiration from physical processes seen in nature,  and those in this category that mimic the evolution of quantum systems are known as \emph{quantum-inspired optimization} methods.  

Quantum-inspired (or ``dequantized'') algorithms have arisen in recent years of out a rich interplay between physics and algorithms research in the context of quantum computing. 
Thus, as notions of complexity now find analogues in many-body systems, so do quantum dynamics inform the design of quantum and classical algorithms. 
In the area of classical optimization, two quantum algorithms have generated considerable interest: quantum annealing and the quantum approximate optimization algorithm (QAOA). 
Promising developments in quantum annealing have inspired classical heuristic algorithms such as simulated quantum annealing~\cite{crosson2016} and sub-stochastic monte carlo~\cite{jarret16}, both of which mimic the evolution of the quantum state under an adiabatically evolving Hamiltonian.  

Recent results on the performance of shallow-depth QAOA on the problem of \textsc{max-e3-lin2} led to improved approximations of corresponding classical algorithms for the same problem~\cite{farhi2014,barak2015}. 
More recently, a new classical heuristic known as Local Tensor (\lt{}) was introduced in \cite{hastings2019}, taking inspiration from QAOA and closely related to previously known classical heuristics for distributed computing~\cite{hirvonen2014}. 
It was shown in \cite{hastings2019} that \lt{} has average-case performance better than single-layer QAOA for triangle-free \maxc{} and \textsc{max-k-lin2} by tuning only one global hyperparameter, in contrast to the two-parameter tuning required for QAOA. 
Currently, it is unknown whether \lt{} may be useful as a heuristic more broadly, and if so, how the hyperparameters should be set in practice. 
In this paper, we address this question by implementing a version of \lt{} and benchmarking it on the problem of \maxc{}. 
We find that on the instances studied, the performance of \lt{} can be considerably enhanced by hyperparameter tuning, and that it is possible to provide good initial guesses on the hyperparameters as function of the instance description. 
Under such settings, the performance of \lt{} is comparable to the performance of the commercially available solver, Gurobi. 

Our setup is described in~\cref{sec:SpinProblems,sec:LT,sec:Instances}, followed by a discussion of hyperparameter tuning and the underlying physics of the solver in~\cref{sec:Tuning,sec:LTDependence}, respectively. Then, we compare the performance of tuned \lt{} with those of Gurobi (\cref{sec:LTvsGurobi}) and gradient descent (\cref{sec:LTvsGD}). Finally,~\cref{sec:ImaginaryTime} discusses the similarity of our implementation of \lt{} to discrete, imaginary-time Schr\"odinger evolution of the spin system under the problem Hamiltonian.

\section{Spin problems}
\label{sec:SpinProblems}
We refer to binary unconstrained optimization problems on spin degrees of freedom $s_i\in\curly{-1,1}$ as spin problems. 
Most generally, one can express the objective function as a polynomial in the variables. 
Furthermore, since higher powers of the binary variables are trivial, the polynomial is guaranteed to be degree at most one in each variable. 
The objective function to be maximized can therefore be viewed (up to a negative sign) as a Hamiltonian of a spin system, and the optimization problem maps to sampling from the ground state of the Hamiltonian. 
The cost Hamiltonian for a system of $n$ spins with indices $\curly{1,2,\ldots,n}$ can be written as
\begin{equation}
\label{eq:CSPpolynomial}
  H = \suml{\alpha}{}{w_{\alpha}\prodl{i\in\alpha}{}{s_i}}.
\end{equation}
Therefore, $H$ is a sum of monomials, or \emph{clauses}, where a clause $\alpha$ is supported on the subset of spins $\alpha\subseteq \curly{1,\ldots, n}$. 
The sum is weighted by clause weights $w_\alpha$. The problem can be fully specified as a weighted hypergraph $G = (V,E,W)$, on vertices $V = \curly{1,2,\ldots,n}$, hyperedges $E = \curly{\alpha,\cdots}$, and clause weights $W = \curly{w_\alpha, \ldots}$.

A wide range of optimization problems be cast as binary unconstrained maximization problems~\cite{lucas14}, making this problem description very versatile. 
A simple (and commonly studied) case is one where the polynomial~\eqref{eq:CSPpolynomial} is quadratic, i.e. $|\alpha|\le 2$ for every $\alpha$. 
This case captures several interesting physical systems such as Ising spin glasses, as well as a wide range of graph optimization problems. 
In this work, we focus on a particular quadratic spin problem, \maxc{}, defined in the following manner. 
Given a weighted graph $G=(V,E,W)$, we define a \emph{cut} to be a partition of the vertices of the graph into two sets. 
The weight of the cut (or simply the cut) is then defined as the sum of weights of edges going across the cut. 
Therefore, for any $A\subset V$, the cut is
\begin{equation}
  F(A) := \suml{i\in A, j\in \bar{A}}{}{w_{ij}}.
\end{equation}
Then, given a graph $G$, \maxc{} asks for the largest cut of the graph. 
To show that \maxc{} can be written as a quadratic spin problem, we consider the following encoding: Assign a spin $s_i$ to vertex $i$. 
Then, there is a one-to-one mapping between bipartitions of $V$, $(A,\bar A)$, and spin configuration, $\vec{s} = \paren{s_1,s_2,\ldots, s_n}$, namely, by setting $s_i=+1$ if $i\in A$ and $-1$ otherwise.
Edges $ij$ that lie wholly in either $A$ or $\bar A$ do not count towards the cut, while edges between $A$ and $\bar A$ do.
In terms of the spins, edge $ij$ will count towards the cut iff the spins $s_i,s_j$ have opposite sign. 
Therefore, we may express the \maxc{} Hamiltonian in the following manner:
\begin{align}
\label{eq:maxcutH}
  H_{\maxc{}} &= \frac{1}{4}\suml{i,j}{}{w_{ij}\cdot(s_is_j-1)}.\\
  \label{eq:sJs}
  &\equiv \frac{1}{2}\vec{s}^T\cdot J\cdot\vec{s}
\end{align}
where $J_{ij} := w_{ij}/2$ with zero diagonal terms, $J_{ii}=0$, and the last equivalence is an equality up to a constant offset $-\frac{1}{4}\suml{i,j}{}{w_{ij}}$.
Notice that the ground state of $H_{\maxc{}}$ corresponds to the largest cut in $G$. 

Despite its simple statement (and apparent similarity to the polynomial-time solvable problem of \textsc{mincut}), \maxc{} is known to be NP-hard~\cite{karp1972}. 
In fact, assuming the unique games conjecture holds, approximating \maxc{} to within a fraction 0.878.. is NP-hard.
This is also the best known performance guarantee, achieved by the exact classical algorithm due to Goemans and Williamson (GW) on \maxc{} with non-negative weights. 
Custom solvers for \maxc{} that that improve on practical performance while sometimes preserving optimality are known~\cite{burer2002, liers2004, rendl2010}. 

\maxc{} is a well-studied problem and often used as a benchmark for new classical, quantum, and quantum-inspired solvers.
Benchmarking of certain quantum-inspired optimization methods such as the coherent Ising machine~\cite{wang13b} and the unified framework for optimization or UFO (see, e.g.,~\cite{mandra16}) has yielded promising results. 
In the following section, we discuss the \lt{} heuristic framework and set up our implementation of the algorithm. 


\section{Local Tensor framework}
\label{sec:LT}

Before describing our implementation, we review the local tensor (\lt{}) algorithm framework laid out in \cite{hastings2019}. 
The \lt{} framework provides a general prescription for a class of \emph{local} algorithms for the optimization of a Hamiltonian on spin variables. 
In a local algorithm, the state (e.g., a spin configuration) is encoded into the nodes of a graph, and the update rule at every node is local in the graph structure, depending only on nodes that are at most a bounded distance away.
Local state updates therefore require information transfer among small neighborhoods and not the entire graph. 
If the graph has bounded degree, this can provide polynomial savings in the running cost of the algorithm. Additional speedup can be obtained in a true distributed model of computing where each node is an individual processor, and communication among nodes is slow compared to the internal operations of each processor. 

\lt{} is a local algorithm framework for optimization problems on spin degrees of freedom (such as \maxc{}). 
In \lt{}, we first relax the domain of every spin variable from the binary set $\curly{-1,1}$ to a continuous superset such as the real interval $[-1,1]$. 
By convention, we denote \emph{soft} spins (i.e. those in the continuous domain) by letters $u,v$, etc. and hard spins by letters $r,s$, etc. 
Then, \lt{} simulates dynamics of a soft spin vector $\vec v$ in discrete time steps, and, at the end of a total number of steps $p$, retrieves a hard spin configuration $\vec{s}$ from the final state via a rounding procedure applied to the soft spins. 
There is considerable flexibility in this setup, and for ease of study, we construct a specific instance of \lt{} here. 

Suppose we are given a \maxc{} instance whose corresponding Hamiltonian (as in \eqref{eq:sJs}) is $H$. 
Denote the state of spin $i$ at time $t$ by $v_{i,t}$, and the full state vector by $\vec{v}_t = \paren{v_{1,t}, v_{2,t},\ldots, v_{N,t}}$. 
Then, we perform the following steps in order, simultaneously for all spins $i=1,\ldots, N$.   
\begin{enumerate}
    \item Initialize all spins uniformly at random, $v_{i,0}\in [-1,1]$.
    \item For $t=0,1,\ldots,p-1$, update $v_{i,t} \mapsto v_{i,t+1}$ as below:
    \begin{enumerate}
    \item $v_{i,t.5} =  v_{i,t} + cF_{i,t}$ where $F_{i,t} := -\partial H/\partial v_{i,t}$ and $c$ is a real constant.
    \item $v_{i,t+1} = \tanh(\beta v_{i,t.5})$, where $\beta$ is a positive constant.
    \end{enumerate}

    \item After $p$ rounds, round each spin to its sign, $v_{i,p}\mapsto \bar{s}_i = \mathrm{sgn}\paren{v_{i,p}}\in\curly{-1,1}$. Return $\bar{s}_i$.
\end{enumerate}
The final configuration $\vec{\bar{s}}$ is a feasible solution candidate. 
As there the initial configuration $\vec{v}_0$ is sampled at random, an outer loop carries out several independent runs of the algorithm and selects the best solution. 

For an instance of size $n$, the domain of feasible solutions corresponds to the vertices of an $n$-dimensional hypercube. 
The relaxation in \lt{} extends the domain to the full hypercube, which allows for small, incremental updates and a well-behaved cost function, at the cost of making the search space infinite. 
However, the rounding step at the end of the algorithm offsets this drawback in the form of a lenient rounding rule: Return the nearest vertex of the hypercube. 
Therefore, the final state of the graph is only required to lie in the correct quadrant (or $2^n$-ant, to be precise) in order to produce the optimal solution. 

The spin update sequence is carried out for a total of $p$ rounds, each consisting of two steps. 
The \emph{force} $F_i = \partial H/\partial v_i$, calculated for each spin, displaces the spin by an amount proportional to it. 
We refer to the constant of proportionality $c$ as the \emph{response}. 
Next, we apply the nonlinear function $\tanh(\beta v)$ to the spin, with a rescaling factor $\beta$ which we will call the inverse temperature. 
The number of rounds $p$, response $c$, and the inverse temperature $\beta$ form the hyperparameters of the algorithm, which must be fixed (ideally by optimization) before the algorithm is run on an instance. 
In theory, the factors $c,\beta$ can also be made to vary by round under a predetermined or adaptive schedule, in a manner similar to simulated annealing.
Here, however, we will consider them to be constant in time. 
\section{The instances}
\label{sec:Instances}
Several open-access repositories for \maxc{} benchmarking instances are available online. 
In this work, we use the ``Biq Mac'' library~\cite{biqmac}. 
The \maxc{} instances provided here are random graphs with edge weights drawn from a certain probability distribution. 
Instances are categorized by the number of variables $n$, edge density $d$ (i.e. the expected number of non-zero weight edges) and the edge weight distribution used. 

We organize these benchmarking instances in ~\cref{tab:results}.
\begin{widetext}

\begin{table}[H]
  \centering
  \begin{tabular}{ccccc}
    \toprule[0.5pt]
    \head{Problem type} &\head{Weight distribution} & \head{$n$} & \head{$d$} & \head{Other}\\
    \midrule[0.5pt]
    \texttt{g05\_$n$} & $w_{ij}\in\curly{0,1}$ & $60, 80,100$ & $0.5$ & --\\ 
    \texttt{pm1s\_$n$} & $w_{ij}\in\curly{-1,0,1}$& $80,100$ & $0.1$ & --\\ 
    \texttt{pm1d\_$n$} & $w_{ij}\in\curly{-1,0,1}$ & $80,100$ & $0.5$ & --\\ 
    \texttt{w$d$\_$n$} & $w_{ij}\in\brac{-10,10}$ & $100$ & $0.1, 0.5, 0.9$ & --\\ 
    \texttt{pw$d$\_$n$} & $w_{ij}\in\brac{0,10}$ & $100$ & $0.1, 0.5, 0.9$ & --\\ 
    \texttt{ising$\sigma$-\_$n$} & $w_{ij}\propto \frac{\epsilon_{ij}}{|j-i|^\sigma}$, $\epsilon_{ij}\sim\mathcal{N}(0,1)$ & $100,150,200,250,300$ & -- & $\sigma=2.5,3.0$ \\ 
    \texttt{t$D$g$n^{1/D}$} & $D$-dim. toroidal grid, $w_{<ij>}\in\curly{-1,1}$ & $5^D, 6^D, 7^D$ & $\frac{2D}{n-1}$ & $D=2,3$\\ 
        
        \bottomrule[0.5pt]
  \end{tabular}
  \caption{The benchmarking problems. Each instance is a random graph on $n$ vertices whose edge weights are chosen from the given distribution. 
  In many cases, the overall density of clauses $d$, or, the ratio of non-zero weight edges, is fixed. 
  The last two instances are random spin models borrowed from physics. 
  The \texttt{ising} instances are a 1-dimensional Ising model with long-ranged interactions falling off as a power $\sigma$ of the inter-spin distance, with a numerator randomly sampled from the normal distribution. 
  Finally, the \texttt{t$D$g} instances are periodic, $D$-dimensional spin lattices with random couplings $\pm 1$ along the edges of the lattice.}
\label{tab:results}
\end{table}

\end{widetext}


\section{Hyperparameter optimization} 
\label{sec:Tuning}
In order to talk about the performance of $\lt{}$ on any given instance, we must first consider variations in performance due to parameter setting and randomness. \lt{} (as implemented here) is a family of algorithms in the hyperparameters $c,\beta, p$. Moreover, for fixed hyperparameters, any run of the algorithm has randomness due to the choice of initial spin configuration. Therefore, the energy output at the end of a single run of \lt{} is a random variable dependent on $\paren{c,\beta,p}$.
The median final energy with fixed hyperparameters, however, is a determinate quantity, which we denote $E_{c,\beta,p}$. 

\begin{equation}
   E_{c,\beta,p} = \mathrm{median }_{\vec v_0\in \brac{-1,1}^{\times n}} \lt{}_{c,\beta,p}(\vec v_0)
\end{equation}
where, by abuse of notation, $\lt{}(\vec v)$ denotes the output energy of \lt{} with input configuration $\vec v$. 
By definition, half of the runs of LT are expected to produce an optimum with energy lower than $E$, making the median energy a useful figure of merit. 
The true median energy can be approximated in practice by the median value of $M$ independent runs of $\lt{}_{c,\beta,p}$,  
\begin{equation}
   \mathrm{median }\curly{E_1,E_2,\ldots, E_M} = \tilde{E} \approx E_{c,\beta,p}
\end{equation}
Since we ultimately wish to study the performance of LT as a whole, the hyperparameters must be fixed via a well-defined procedure that takes as input the instance description and returns an (ideally optimal) hyperparameter setting. 
The most rigorous criterion is global optimization of the performance with respect to each hyperparameter independently. This is important, e.g. to avoid spurious trends in the runtime scaling that arise from imperfect hyperoptimization. 

Since this is a computationally expensive task, we focus first on gaining a better understanding of the effect of the hyperparameter on the algorithm performance, and construct metaheuristics to minimize the resources needed for hyperparameter optimization.   

\begin{figure}[tb]
    \centering
    \includegraphics[clip=true, trim = 0pt 0pt 35pt 30pt, width=\linewidth]{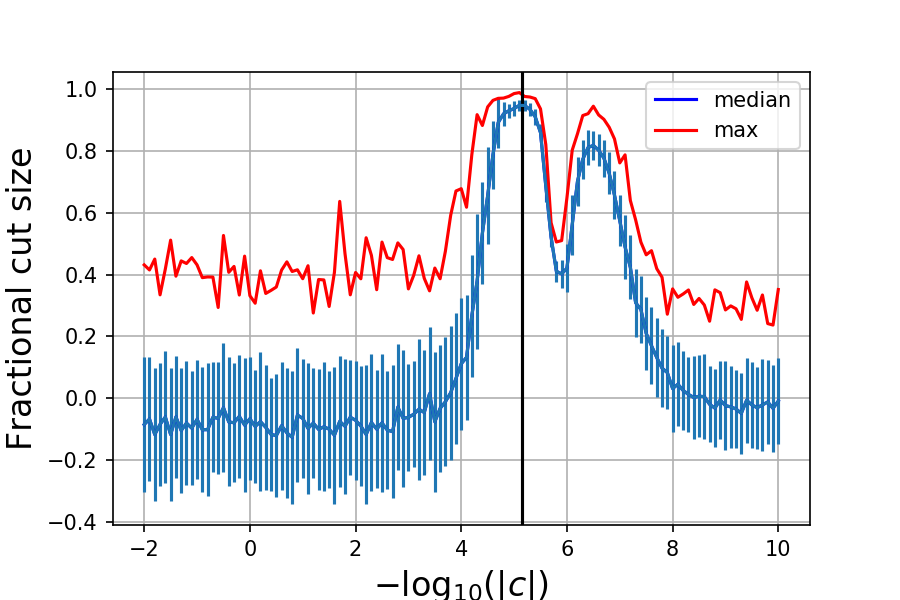}
    \caption{Hyperparameter sweep for the instance \texttt{ising2.5-100}. We vary $c$ over multiple orders of magnitude, and plot the median and max value of optimum found for several runs of the algorithm. It can be seen that peak performance is found when $c\sim \bar c $ (indicated by the black line).}
    \label{fig:c-beta-tuning}
\end{figure}

\subsection*{Response $c$.}
The response $c$ is the sensitivity of the spins to force. 
Intuitively, setting $c$ too large or too small would make the spins too responsive to displacement or frozen, respectively. 

Therefore, we expect a regime for $c$ values where the spins are optimally sensitive to the force, and the algorithm should also perform well in this regime. 
Given a typical length of spins $v_i\sim 1$ and maximum possible force on spin $i$, $F_i\sim \suml{j=1}{N}{|J_{ij}|}$, a natural guess for $c$ is the inverse of the maximum force. We define
\begin{equation}
\label{eq:barc}
\bar c =  2\left\langle\suml{j=1}{N}{|J_{ij}|}\right\rangle_i^{-1}
\end{equation}
where the brackets $\langle\cdot\rangle_i$ denote a mean over all sites in the graph. The factor of two is chosen purely empirically.
As shown in~\cref{fig:c-beta-tuning}, we find that $\bar c$ is indeed a natural scale for the response, and optimal performance is typically found to be within an order 1 factor of $\bar c$. 
Hereafter, we use a rescaled hyperparameter $\eta := c/\bar c$. 
\begin{figure}[tb]
    \centering
    \includegraphics[trim = 10pt 10pt 35pt 30pt, width=\linewidth]{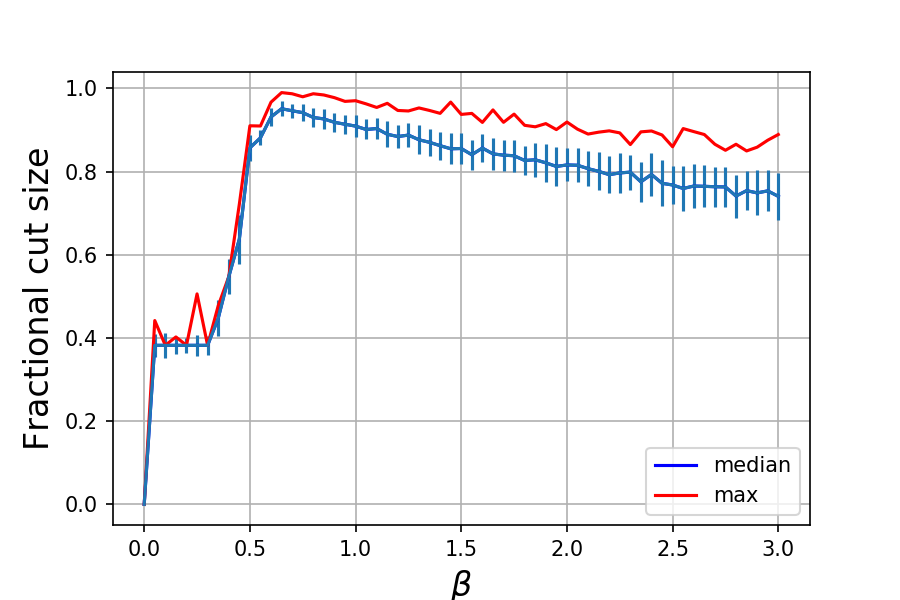}
    \caption{Hyperparameter sweep for the instance \texttt{ising2.5-100}. We plot the median (blue) and max (red) performance as a function of $\beta$, for a fixed value of $c\sim \bar {c}$. The performance is sensitive to order 1 variation in $\beta$, varying from sub-random (cut fraction $0.5$) to close to optimal at $\beta \simeq 0.7$. This behavior is typical across all instances studied.}
    \label{fig:beta-tuning}
\end{figure}

\subsection*{Inverse temperature $\beta$.}
The $\beta$ parameter scales the value of the input to the $\tanh$ activation function.  
Intuitively, this enables mapping the displaced spin to the linear response region of the $\tanh$ function for maximum sensitivity. 
This also ensures that the spin stays of order 1 and therefore sensitive to forces applied in subsequent rounds. 

\begin{figure}[tb]
    \centering
    \includegraphics[clip=true, trim = 10pt 0pt 45pt 40pt,width=\linewidth]{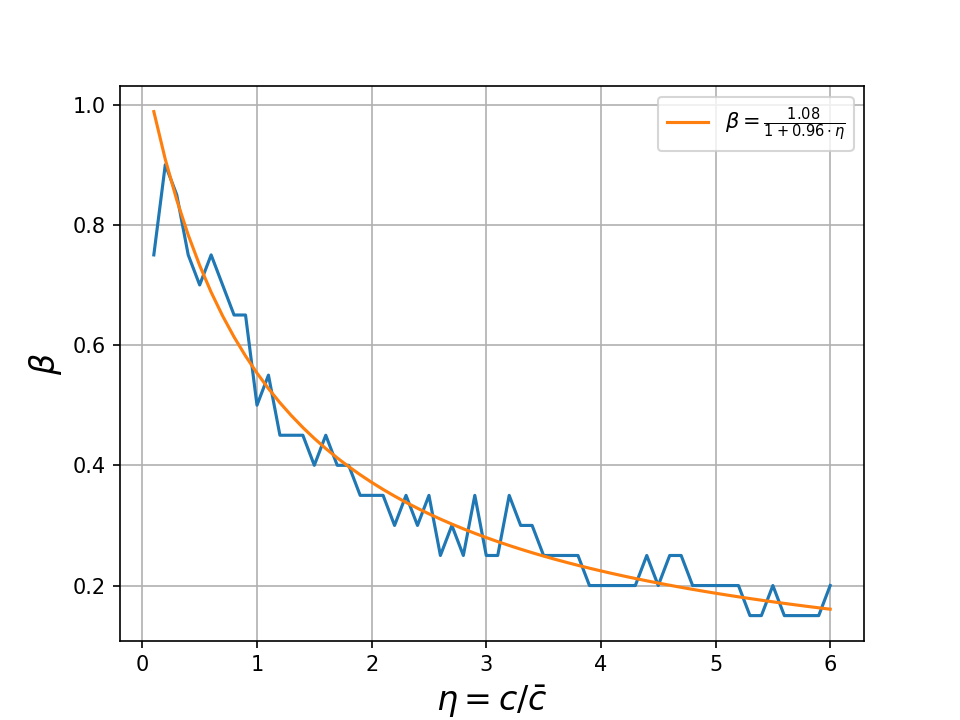}
    \caption{Optimal $\beta$ for a range of $\eta$ values, plotted for a torus instance. For each $\eta$, the optimal $\beta$ was found by grid search. The fit to the functional form given in~\eqref{eq:BetaEta} is also shown. The curve profile and quality of fit seen here are typical to all instances studied.}
    \label{fig:BetaEtaFits}
\end{figure}

\begin{figure}[tb]
    \centering
    \includegraphics[clip=true, trim = 19pt 5pt 40pt 40pt, width=\linewidth]{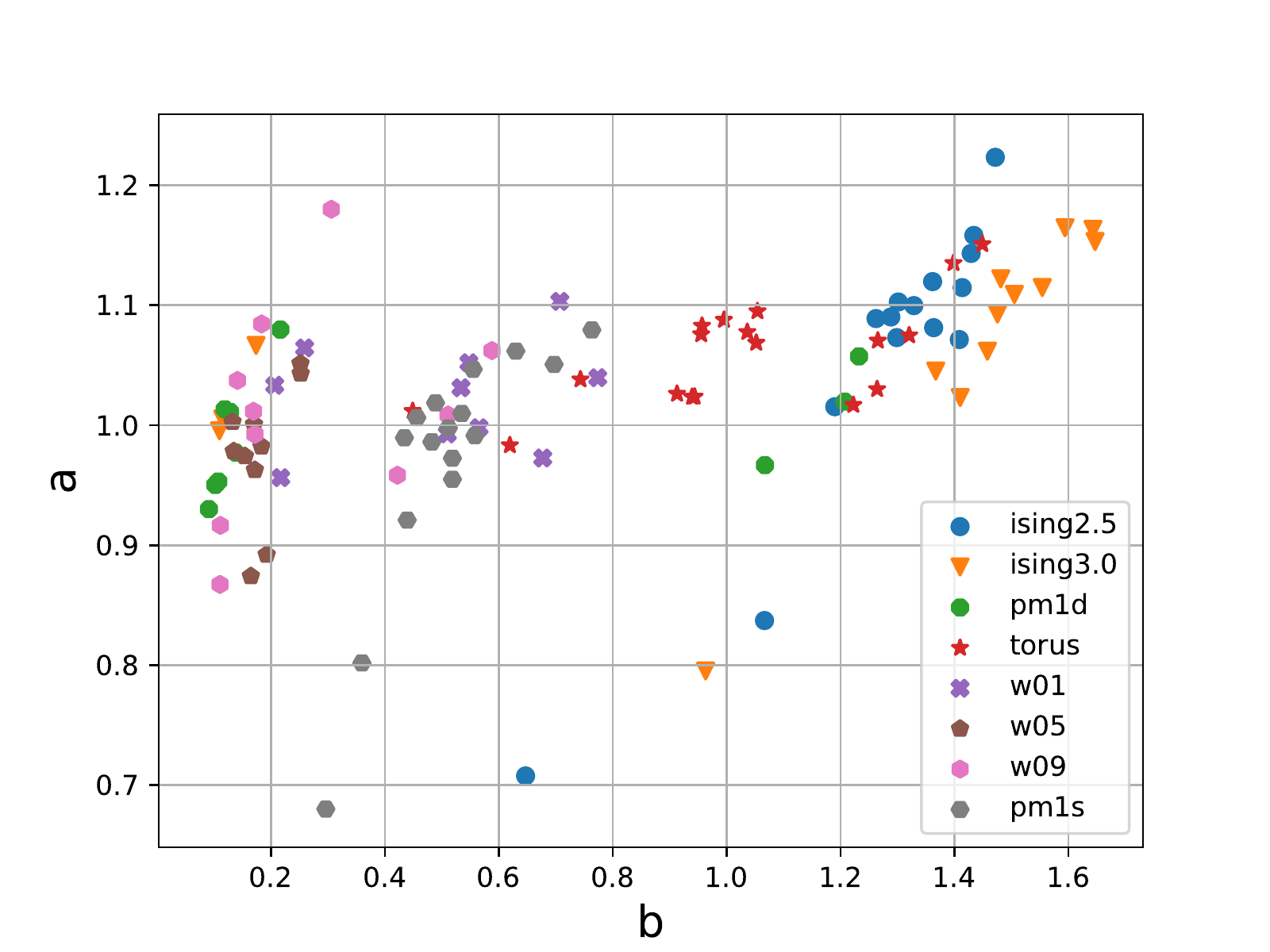}
    \caption{Clustering in the fit coefficients $a,b$ in~\eqref{eq:BetaEta} for different instances (see \cref{tab:results}). 
    We see that the fitting numerator $a$ is close to 1 for most instances, while $b$ varies considerably. There is a reasonable degree of clustering by instance type in $b$.}
    \label{fig:BetaEtaClustering}
\end{figure}

Just like the response, we can now make an educated guess for $\beta$. 
For a fixed response $\eta = c/\bar{c}$, spin $v_i$ is displaced as $v_i\rightarrow \tanh{\beta (v_i + cF_i)}$. 
Since $v_i\leq 1$ for all $i$, the argument of the tanh function must lie between $[-\beta(1+\eta),+\beta(1+\eta)]$. 
In order to be maximally sensitive to displacement, we should set $\beta$ such that $\beta(1+\eta) \sim O(1)$. 
This gives us a functional dependence between $\beta$ and $\eta$ as 
\begin{equation}
\label{eq:BetaEta}
    \beta = \frac{a}{1+b\eta},
\end{equation}
where we have introduced two fitting parameters $a,b$. 
From the available instances, this relationship can be checked by extracting the locus of optimal settings for $(\eta,\beta)$ and fitting them to the above functional form. 
The results are shown in~\cref{fig:BetaEtaFits}. 
The quality of fits suggests that the functional dependence given in ~\cref{eq:BetaEta} is accurate. 
In~\cref{fig:BetaEtaClustering}, we show how the coefficients $a,b$ cluster for different problems. 

From this analysis, we see that given a problem instance, the hyperparameters $\eta,\beta$ can be guessed with very little optimization, and tuned further, if necessary, by local search in a range of order 1 in each parameter. 

\subsection*{Number of rounds $p$.}
 The number of rounds $p$ required by the algorithm is dictated by the convergence to steady state. Qualitatively, this may be connected to the rate of information propagation in the graph, via quantities such as the girth. Unlike for $c,\beta$ however, a direct guess for $p$ may be harder to obtain. 

Instead, we use a dynamic criterion to set the value of $p$. 
Since \lt{} is iterative and closely related over gradient descent, we expect that at some point during the algorithm, the spin vector attains a steady state such that all subsequent displacements are smaller than a given threshold. 
As the final state is determined by the quadrant containing the vector and not the exact value of the vector, small displacements have a small or no effect on the outcome. 

This convergence in the state vector is seen across different instances. 
Therefore, once the displacement of the state falls below a set threshold, we terminate the algorithm. 
While the threshold is an additional parameter, it can be set to be sufficiently smaller than the size of the hypercube.  

\begin{figure}[t]
    \centering
    \includegraphics[clip=true, trim =0pt 0pt 30pt 10pt, width=\linewidth]{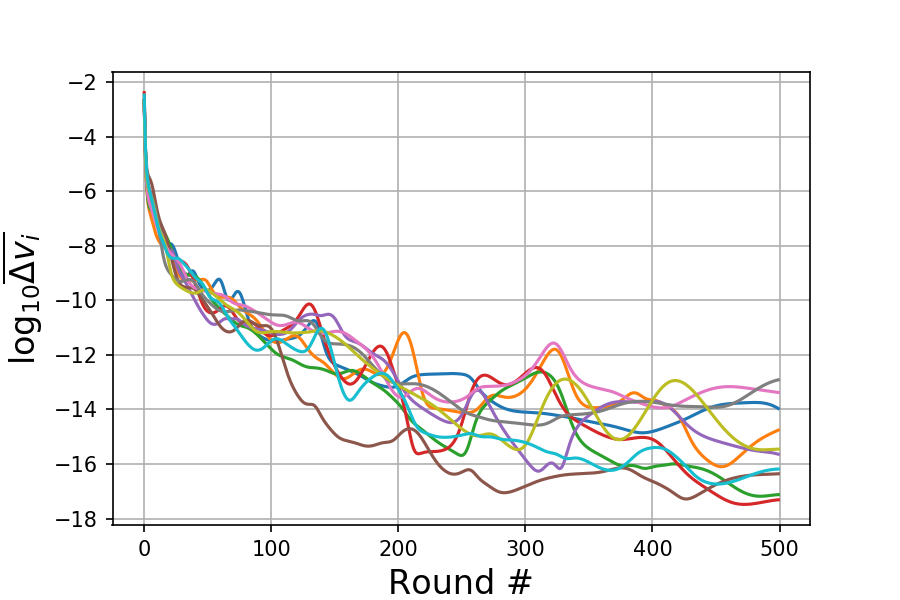}
    \caption{Displacement in the spin variables between successive rounds, as a function of round number. 
    We have chosen ten spins from a 100-spin instance of type \texttt{w05} for illustration. 
    The log displacement becomes small after a short number of rounds, indicating that the updates can be terminated early for the rounding step.}
    \label{fig:pConvergence}
\end{figure}

\section{Dependence of \lt{} dynamics on the hyperparameters}
\label{sec:LTDependence}
The implementation of \lt{} studied here allows three hyperparameters, namely, the number of rounds $p$, the response to force $c$, and the ``inverse temperature" $\beta$. We have discussed how to set these hyperparameters for a given \maxc{} instance, giving (in the case of $c,\beta$) good initial approximations that depend on the instance description, or (in the case of $p$) a dynamic criterion based on the convergence of the state vector. Here we give a physical description of the system dynamics and show, qualitatively, why it is reasonable to expect such behaviors. 

We will analyze the behaviour of \lt{} near a steady state solution $\bar{v}$ that satisfies $\bar{v}_i = \tanh\brac{\beta\paren{\bar{v}_i + c\bar{F}_i}}$ for all spins $i$. Note that a steady state always exists: the all-zero state $\bar{v}_i=0$ is an example. More generally, the transcendental equation for steady state, while not guaranteed to have other solutions, can be approximated as a linear equation when $\bar{v}_i\ll 1$, which has non-zero solutions for particular values of $\beta, c$. Generically, we expect other steady solutions lying within the hypercube, and find this to be true in our numerics. 

\begin{figure}[tb]
    \centering
\includegraphics[clip=true, trim = 5pt 0pt 40pt 40pt, width=\linewidth]{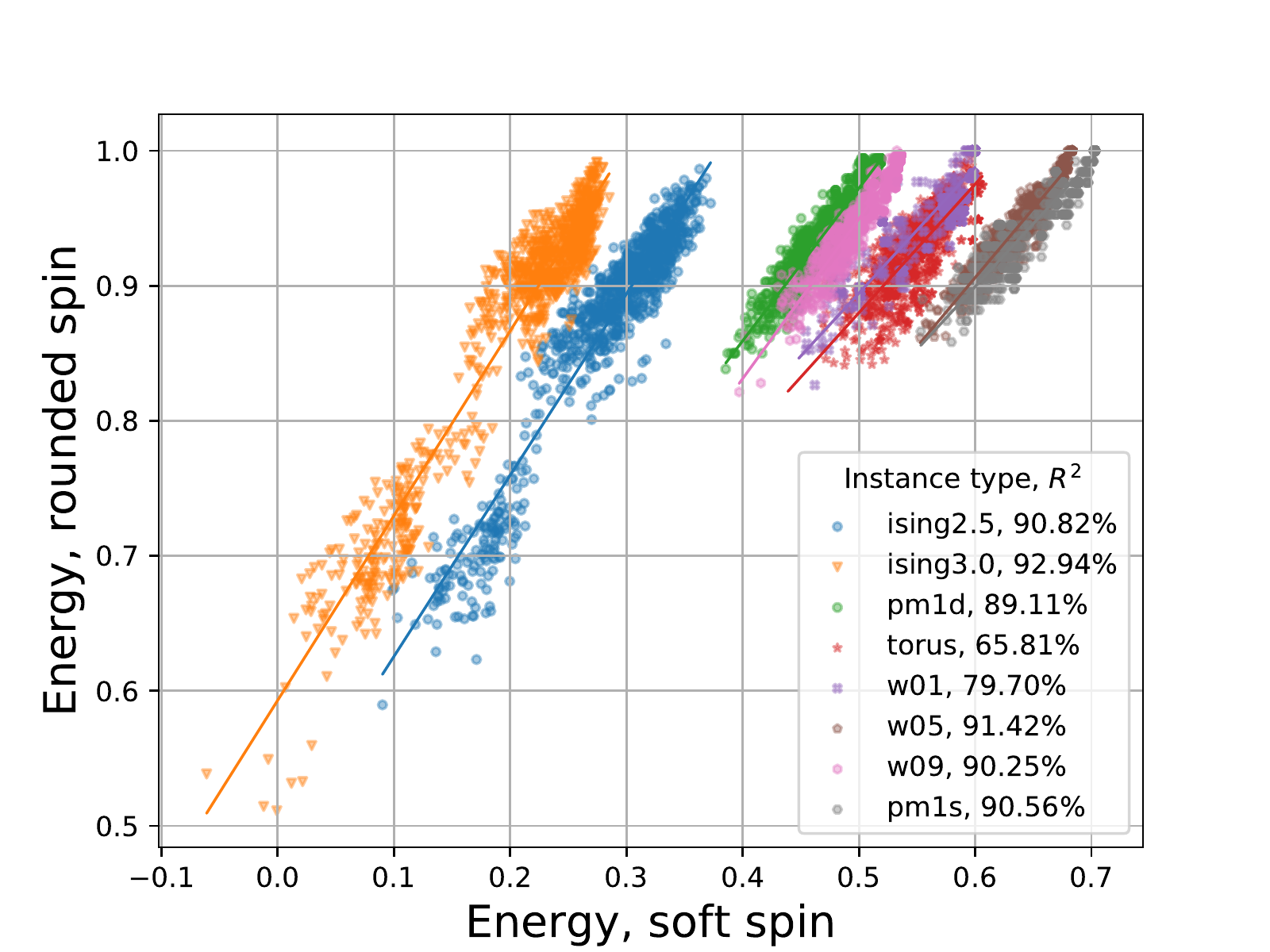}    
    \caption{We plot the objective function evaluated at the soft spin (y axis) and the corresponding rounded spin obtained at the end of an \lt{} run, for several independent runs of the algorithm on eight instances. Each instance is picked from a different instance type. The correlation between the two quantities (given by $R^2$ values in the legend) indicates that better steady state soft spin configurations tend to map to better feasible solutions.}
    \label{fig:rounding}
\end{figure}

Suppose, for a given run of the algorithm, the system tends to a particular steady state $\bar {v}$ at long times, with the state at some finite time $t$ given by $\vec{v}_t = \vec{\bar{v}} + \vec{\delta}_t$, where $|\delta_{i,t}|\ll |\bar{v}_i|$. Then, to first order in the displacement, we have 
\begin{align}
    v_{i,t+1} &= \tanh\brac{\beta\brac{\paren{\id +cJ}\cdot\paren{\vec{\bar{v}} + \vec{\delta_t}}}_i}\\ &=  \tanh\brac{\beta\paren{\bar{v}_i + c\bar{F}_i} + \beta\brac{\paren{\id +cJ}\cdot\vec\delta}_i}\\
    &\simeq \tanh\brac{\beta\paren{\bar{v}_i + c\bar{F}_i}}\\ + &\beta\brac{\paren{\id +cJ}\cdot\vec\delta}_i \mathrm{sech}^2\brac{\beta\paren{\bar{v}_i + c\bar{F}_i}}\\
    &= \bar{v_i} + \beta\paren{1-\bar{v}_i^2}\brac{\paren{\id +cJ}\cdot\vec\delta}_i
\end{align}
where we used the steady-state condition, and $\tanh' (x) = 1-\tanh^2(x)$. Therefore, the displacement at time $t+1$ is
\begin{equation}
    \vec{\delta}_{t+1} \simeq \beta\paren{\id-\bar{V}^2}\cdot\paren{\id +cJ}\cdot\vec\delta_t\, ,
\end{equation}
where we defined the diagonal matrix $\bar{V}_{ii} = \bar{v}_i$. Therefore, the norm of the displacement vector close to steady state is bounded as 
\begin{equation}
        |\vec{\delta}_{t+1}| \lesssim  |\beta|\cdot||\paren{\id-\bar{V}^2}||\cdot||\paren{\id +cJ}||\cdot|\vec\delta_t|\, .
\end{equation}
Since $||\paren{\id-\bar{V}^2}||\le 1$, and $||\id + cJ|| \leq 1 + |c|\cdot||J||$, it follows that 
\begin{equation}
        |\vec{\delta}_{t+1}| \lesssim  \beta\cdot\paren{1 +c||J||}\cdot|\vec\delta_t|\, ,
\end{equation}
assuming $c,\beta \ge 0$. Finally, consider the following properties:
\begin{enumerate}
    \item Since $J$ has zero diagonal, then by the Gershgorin circle theorem, all eigenvalues of $J$ lie within a disc of radius $\max_i\suml{j=1}{n}{|J_{ij}|}$.
    \item If $c = \eta \bar{c}$, where $\bar{c} = \max_i\suml{j=1}{n}{|J_{ij}|}$, then $1 + c||J|| \le 1 + \eta$.
\end{enumerate}
Therefore, for a choice $\beta \gtrsim \frac{1}{1+\eta}$, we expect 
\begin{equation}
    |\vec{\delta}_{t+1}| \lesssim |\vec{\delta}_{t}|
\end{equation}
giving the condition for dynamics converging to a steady state. Three observations can be drawn from this:
\begin{enumerate}
    \item $\bar{c}$ provides a natural unit for the response $c$.
    \item For optimal convergence, we expect the dependence between $\beta$ and $\eta = c/\bar{c}$ to be given by $\beta \simeq a/(1+b\eta)$ for some parameters $a,b$.
    \item Under the above circumstances, the trajectory near steady state is stable and follows an exponential convergence towards the steady state solution. Therefore, the algorithm can be ``safely'' terminated when the displacement is under a certain threshold.
\end{enumerate}

These three observations closely match our empirically derived rules for good performance of \lt{}. This indicates that the steady state solutions may also correlate with the locations of good feasible solutions (given by the nearest hypercube vertex). This can be seen in~\cref{fig:rounding}.

\begin{figure}[tb]
    \centering
    \includegraphics[clip=true, trim =10pt 5pt 45pt 40pt, width=\linewidth]{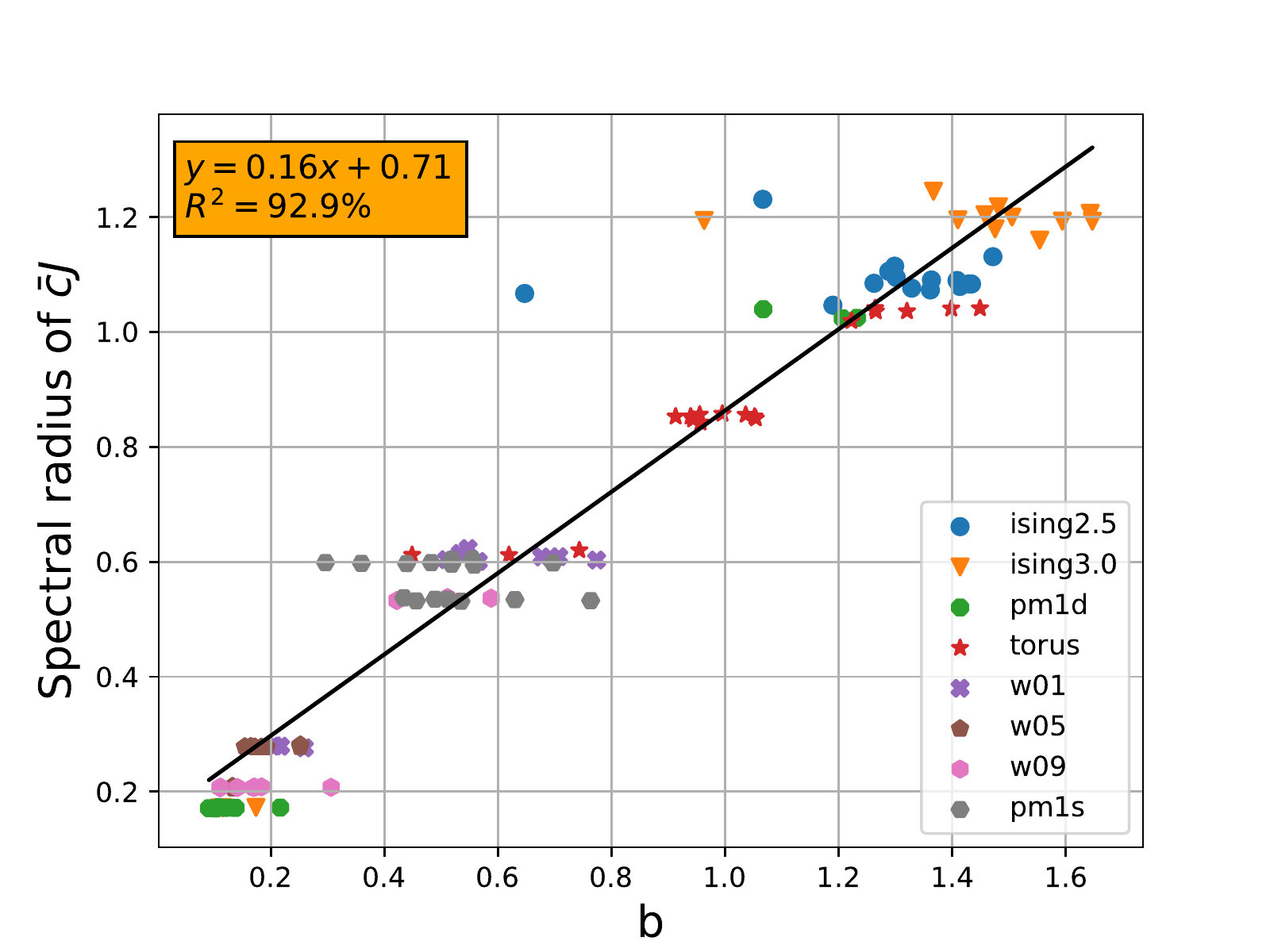}
    \caption{Dependence of the fitting parameter $b$ and the spectral radius of the the normalized coupling matrix $\bar{c}J$. A linear regression fits the data with $R^2=92.9\%$, indicating a strong linear relationship between the two quantities. Therefore, the spectral norm of the coupling matrix can give a good estimate on the fitting parameter $b$ and hence $\beta$.}
    \label{fig:bvs||J||}
\end{figure}

In fact, the quantity $\bar{c}$ is defined not as the maximum but as a mean,~\cref{eq:barc}. However, the functional relationship $\beta = \frac{a}{1+b\eta}$ is still seen to hold for some $a,b$. 

We can study the dependence of $a,b$ by instance type. As shown in \cref{fig:BetaEtaClustering}, the parameters form clusters by instance type, and the value of $a$ is close to 1 for all instances studied. The value of $b$ varies considerably from instance to instance. Looking at the functional form, it is reasonable to guess that $b$ is related to the spectral radius of the coupling matrix $J$. While we bound the magnitude of the largest eigenvalue by $\max_i\suml{j=1}{n}{|J_{ij}|}$, the actual value may be smaller, and $b$ may be understood to reflect this correction. We check this conjecture by plotting the relationship between $||J||$ and $b$ for every problem instance in~\cref{fig:bvs||J||}. 

This relationship can be particularly useful if the largest eigenvalue of the matrix can be calculated or estimated quickly. Then, by inverting the linear regression shown in~\cref{fig:bvs||J||}, one obtains a good initial guess for $b$. The initial guess for $a$, on the other hand, is simply 1. This potentially reduces the hyperparameter optimization to local minimization in the single parameter $\eta$, which is computationally inexpensive.

\section{Comparison with Gurobi}
\label{sec:LTvsGurobi}
Gurobi is a versatile commercial optimization software that solves a broad range of problems including quadratic programming (QP), linear programming, and mixed integer programming.  
Additionally, the software includes in-built heuristics to find good initial solution candidates quickly, as well as ``pre-solve'' subroutines  and simplify the problem description by eliminating redundant variables or constraints. 

In order to use Gurobi, a \maxc{} instance can be relaxed to either a linear program or a quadratic program. 
Mapped to an LP, the instance is specified by real variables $x_{ij}\in\mathbb{R}$ to each edge, with the inequality constraints $x_{ij}\le 1$, where $x_{ij}=1$ if and only if the edge $ij$ is in the cut. 
Additional constraints follow by observing that not all configurations are feasible: for example, for three edges $ij,jk,ki$, at most two may be part of a cut. 
Any feasible solution must satisfy such \emph{cycle} constraints as well, expressible as inequalities of the form $x_{ij} + x_{jk} + x_{ki} \le 2$. 
Then, the LP is formulated as maximization of the objective function $\vec{w}^T \vec{x}$, subject to the above inequalities, where $\vec{w}$ represents the vector of edge weights. 
While the number of cycle inequalities are exponential, it is possible to solve the separation problem in polynomial time~\cite{barahona1986}, leading to a cutting-planes algorithm for finding and including violated constraints dynamically for each successive iteration of the LP. 
However, this formulation is somewhat unnatural for \maxc{}, and is seen to perform poorly for dense graphs, due to a blowup in the number of variables. 

A more natural formulation is as a QP with linear constraints, where every vertex $i$ is assigned to one variable $s_i$, and the problem is expressed as 
\begin{align*}
    &\max~\frac{1}{4}\vec{s}^T\cdot W\cdot\vec{s}\\
    &\mathrm{s. t. }\ -1\leq s_i \leq 1.
\end{align*}
Here, the matrix $W$ is the weighted graph Laplacian, with $w_{ii} = \suml{k\neq i}{}{w_{ik}}$ and $W_{ij}=-w_{ij}$ for $i\neq j$. 


The QP formulation is convex when the edge weights are non-negative, and not otherwise. 
However, the weight matrix $W$ can be made convex trivially by the addition of a suitably chosen diagonal matrix $D$. Since we seek spin configurations that must satisfy $s_i^2=1$ for all spins $i$, the diagonal merely shifts the objective function by a constant everywhere on the feasible solutions. 
For a sufficiently large shift, the matrix becomes positive semi-definite by the Gershgorin circle theorem: All eigenvalues of 
$W+D$ lie in the union of all discs $S_i\in\mathbb{C}$ with centers given by $w_{ii}+D_{ii}$ and radii $\suml{}{}{|w_{ij}|}$. Then $D_{ii}$ can always be chosen large enough so that each disc lies wholly in the right half-plane, so that all eigenvalues of $W-D$ are non-negative. This makes the shifted maximization problem convex, implying that solutions to the relaxed problem yield feasible solutions of the discrete problem. 

Gurobi has support for both convex and non-convex QP, and uses interior point methods (specifically, a parallel barrier method) and the simplex method to solve problems in QP formulation. 
We provide the instances to Gurobi in QP formulation. 
\begin{figure}[h]
    \centering
    \includegraphics[width=\linewidth, clip=true, trim=0pt 0pt 35pt 35pt]{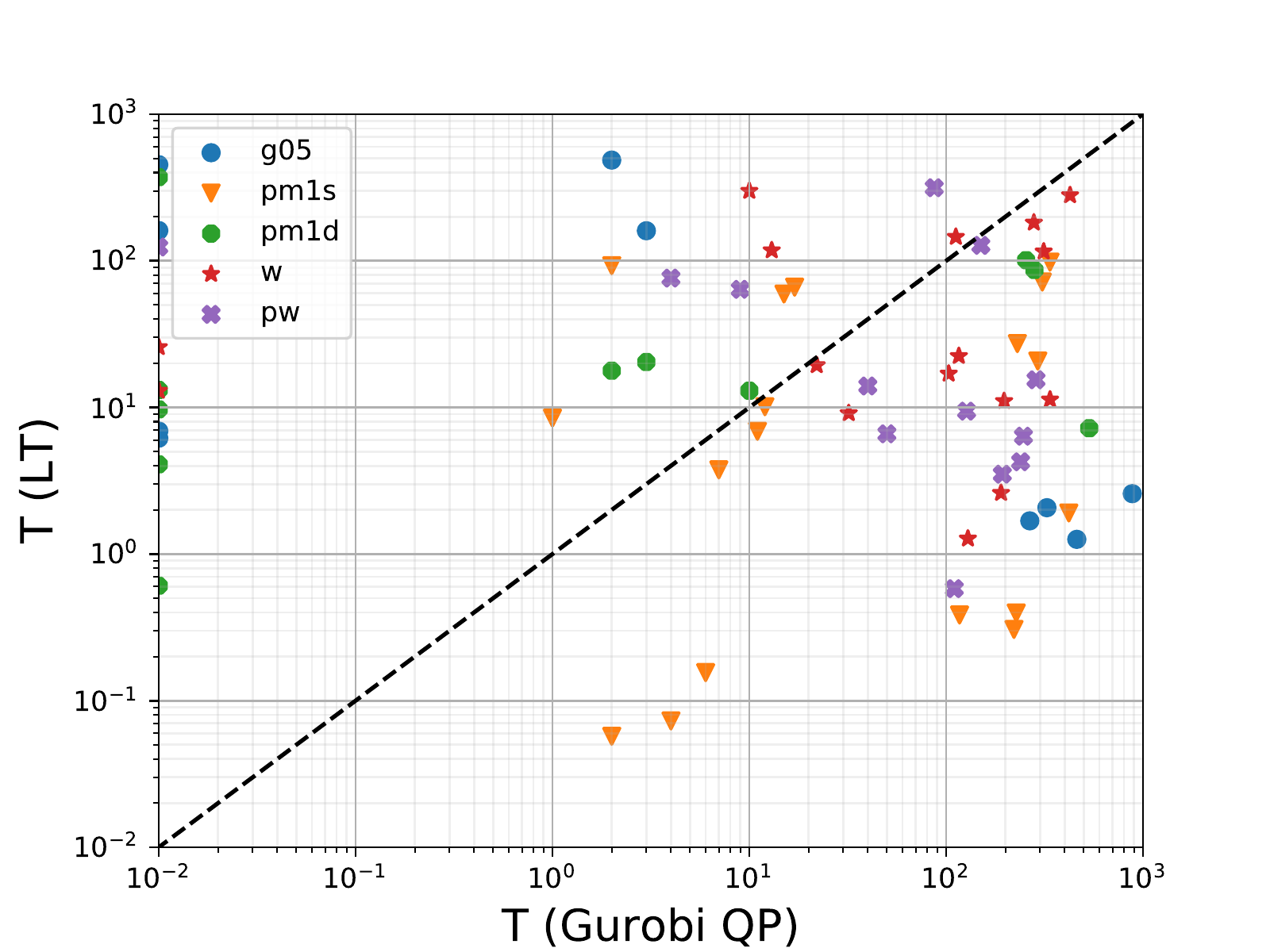}
    \caption{Performance of \lt{} compared against Gurobi for several benchmarking instances. 
    The performance metric used is median time taken to find the optimum (over 10 runs), with a timeout of $10^3$ seconds. Timed out instances are not shown: Out of 130 instances, LT and Gurobi timed out on 47 and 22 instances, respectively, including 7 instances where both timed out. Times faster than a certain threshold are reported by Gurobi as 0s (corresponding to points along the left edge). 
    \lt{} and Gurobi find optima faster than each other in an equal number of instances, with no clear instance-dependent advantange. The speedup on either side is in some cases up to three orders of magnitude.}
    \label{fig:LTvsGurobi}
\end{figure}
Then, we can compare the time to find optimal solution for Gurobi and \lt{} on our benchmarking instances. The time has to be carefully defined in each case for a fair comparison. 
Gurobi is a deterministic algorithm, except for an initial (optional) heuristic step for proposing an initial solution candidate which takes a small fraction of the total runtime. 
The algorithm terminates when the optimum is found and proved. The latter typically requires additional time to improve the upper bound on the optimum until it matches the best optimum found. 
Since we use benchmarking instances that have known optima, we define runtime leniently as the time to find (but not necessarily prove) the optimal solution.  

On the other hand, \lt{} is randomized due to the random choice of initial state, and multiple runs are necessary to gather statistics on the performance. 
Therefore, we define runtime as the median performance over 30 independent runs of \lt{} for each instance. 
Furthermore, since \lt{} requires parameter tuning, we allow up to 20 $s$ of hyperparameter tuning by grid search in $\beta,\eta$ that is not considered part of the runtime. 
Note that the results of~\cref{sec:Tuning} suggest that the parameters can be set automatically, either adaptively as for $p$ or by a well-motivated formula for $\eta,\beta$, without the need for a full grid search. 

Then, as shown in~\cref{fig:LTvsGurobi}, the runtime performance of \lt{} and Gurobi can be compared directly on every instance. 
We see that there is significant spread in performance for every problem type, for both \lt{} and Gurobi. 
Promisingly, there are instances in every problem class for which \lt{} is significantly faster than Gurobi. 

The comparison with Gurobi illustrates that there may be cases where properly tuned LT can outperform state-of-the-art solvers at a fraction of the time cost. It is pertinant to ask whether the success of LT over other solvers can be predicted in advance, using instance data (or quantities derived from it). We briefly address this question. 

The most obvious performance indicator is the number of variables $n$. The instances used in the time comparison with Gurobi were of size 60, 80, or 100. Another elementary indicator is clause density, or the mean number of clauses per variable, which for a weighted instance is the average row sum of the graph adjacency matrix, $m := \suml{i,j}{}{J_{ij}}/n$. We also compute the average row sum of the absolute value of the weight matrix $\bar{m} := \suml{i,j}{}{|J_{ij}|}/n$. Finally, the misfit parameter $\mu$ measures the degree of frustration in the model. More precisely, it is the ratio of the ground state energy of the model to the ground state energy of a frustration-free reference system. For a given \maxc{} instance, a reference system with all weights $J_{ij}$ replaced by their negative absolute values $-|J_{ij}|$ is frustration-free, with a ground state energy of $-\suml{i<j}{}{|J_{ij}|}$. On the other hand, the ground state energy of the original instance is bounded below by $-\suml{i<j}{}{J_{ij}}$. Therefore, we define misfit as
\begin{equation}
  \mu := \frac{\suml{i<j}{}{J_{ij}}}{\suml{i<j}{}{|J_{ij}|}}\, .
  \label{eq:misfit}
\end{equation}
Then, we ask: How well does a given performance indicator predict the runtime of LT (or Gurobi) on a randomly chosen instance? More formally, treating the runtime and indicator as random variables $X, Y$ respectively, the predictive power can be expressed as the conditional entropy $H(Y|X)$, defined as
\begin{equation}
  \label{eq:cent}
  H(Y|X) := -\suml{x\in\mc{X},y\in\mc{Y}}{}{}p(x,y)\log \frac{p(x,y)}{p(x)}\, ,
\end{equation}
where the sum is taken over the support sets of $X,Y$. Informally, $H(Y|X)$ quantifies the number of additional bits needed to specify $X$ given knowledge of $Y$. The largest possible value of $H(Y|X)$ is $\log|\mc{X}|$ for a discrete sample space $\mc{X}$, corresponding in our case to the number of bins used to group the runtimes. We report the conditional entropy normalized by this maximum, so that a normalized entropy of $0$ ($1)$ corresponds to perfect (no) predictability. The results are presented in \cref{tab:cent}. Relative to Gurobi, the performance of LT is marginally more predictable using the instance data. However, clearly discernable relationships between the performance and any of the indicators studied here could not be obtained using the instance data available, suggesting the need for further systematic study.
\begin{table}[tb]
  \centering
  \begin{tabular}{ccccc}
    \toprule[0.5pt]
    \head{Predictor} &&\head{Gurobi} & \head{LT} \\
    \midrule[0.5pt]
    $n$ && 0.73 & 0.69 \\ 
    $m$ && 0.68 & 0.63\\ 
    $\bar m$ && 0.66 & 0.59\\ 
    $\mu$ && 0.56 & 0.53 \\ 
    \bottomrule[0.5pt]
  \end{tabular}
  \caption{A tabulation of the normalized conditional entropy (as defined in \cref{eq:cent}) of different performance predictors with the runtime of Gurobi and \lt{} on the benchmarking instances. Zero indicates perfect prediction, while 1 corresponds to no predictability. The real-valued predictors $m,\bar m, \mu$ were binned into 20 equally spaced intervals, and the runtime was binned into 20 logarithmic intervals spanning the range 0.01s to 1000 s, with an additional bin for timed-out instances ($t>1000$s). }
\label{tab:cent}
\end{table}

\section{Comparison with Gradient Descent}
\label{sec:LTvsGD}
An inspection of the \lt{} implementation reveals that the algorithm is operationally very similar to a gradient descent algorithm. The difference lies only in the fact that we apply a nonlinear $\tanh$ wrapper to each spin value in every step, while gradient descent is fully linear. This raises a natural question: does $\lt{}$ offer any advantage to gradient descent? 

We formalize this comparison. The \maxc{} Hamiltonian does not have a global extremum over $\mathbb{R}^n$, as all of its second (and higher-order) derivatives vanish. 
This implies that a gradient descent algorithm must constrain the state vector to lie within a closed region of $\mathbb{R}^n$; then the optima are guaranteed to lie on the boundary of this region. 
The natural choice of region is the $n$-dimensional hypercube $H_n := [-1,1]^{\times n}$, whose vertices correspond to feasible solutions to the \maxc{} problem. 
Then, any step that displaces the state vector outside $H_n$ must be modified to obey the constraint. 
We implement this by applying a cutoff function to each spin at the end of every displacement step. 
The form of this function is as follows:
\begin{equation}
    \cutoff(x) = \mathrm{sgn}(x)\cdot \min\curly{1,|x|}.
\end{equation}
When applied to each spin as $\cutoff (\beta v_i)$, this function has the effect of projecting every spin component that exceeds an allowed range $[-1/\beta,1/\beta]$ onto the closest boundary of the range. 
The free parameter $\beta$ controls how wide the allowed range should be. 
\begin{figure}[tb]
    \centering
    \includegraphics[clip=true, trim = 10pt 0pt 35pt 40pt, width=0.5\linewidth]{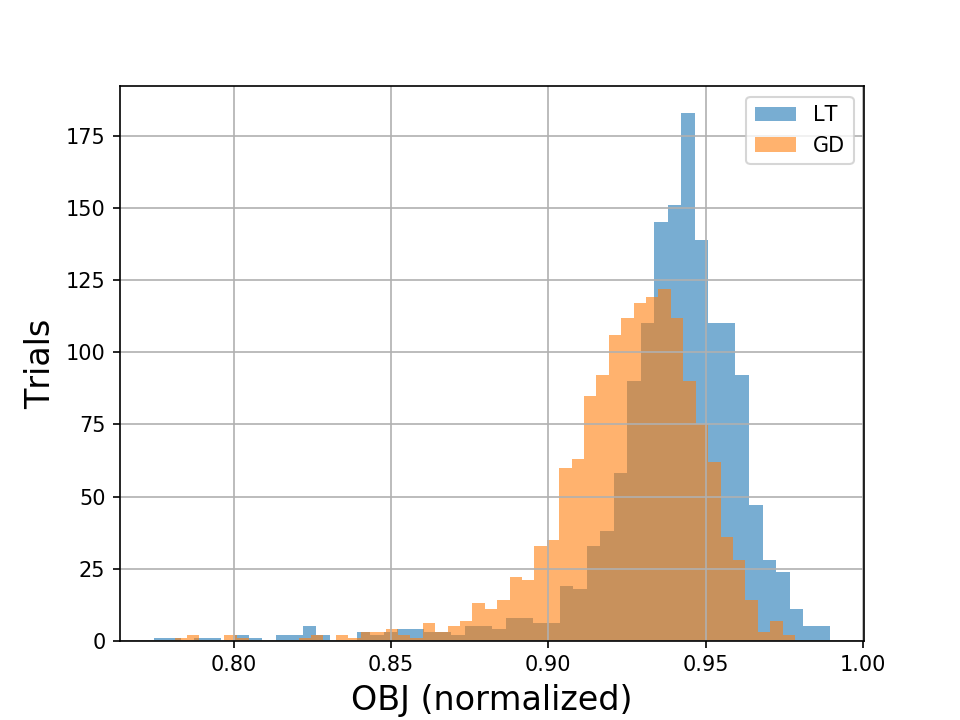}\includegraphics[clip=true, trim = 10pt 0pt 35pt 40pt, width=0.5\linewidth]{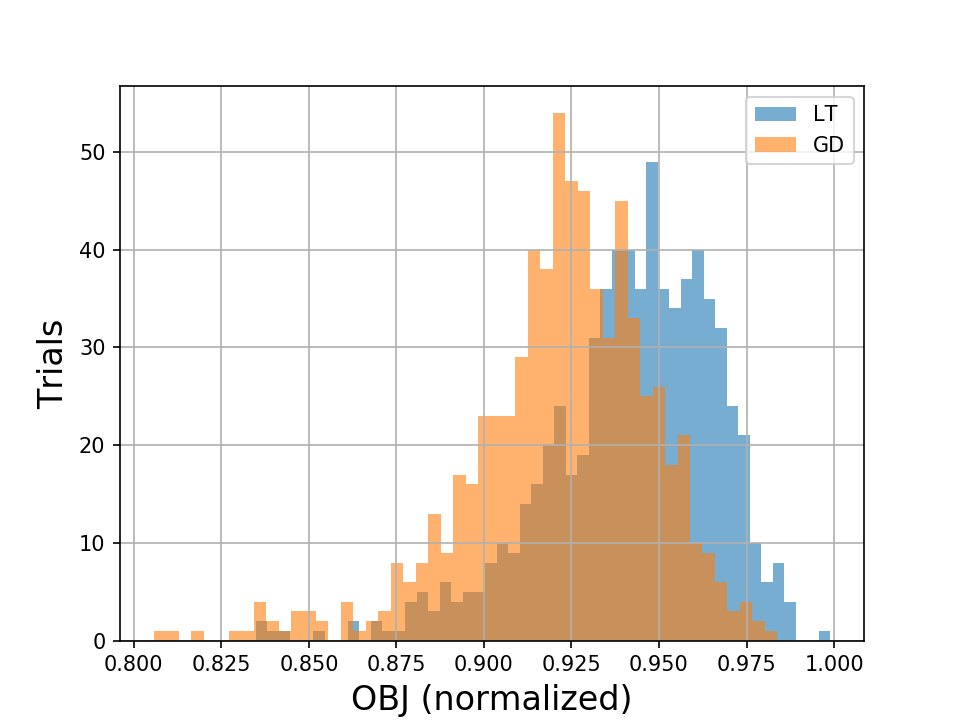}
    
    \includegraphics[clip=true, trim = 10pt 0pt 35pt 40pt,width=0.5\linewidth]{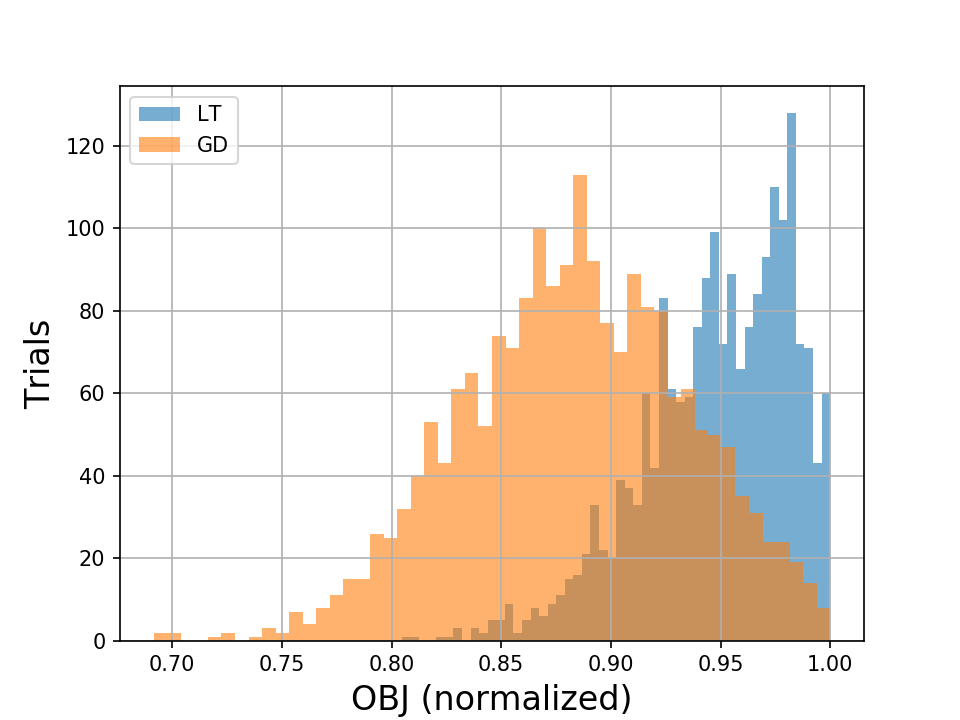}\includegraphics[clip=true, trim = 10pt 0pt 35pt 40pt,width=0.5\linewidth]{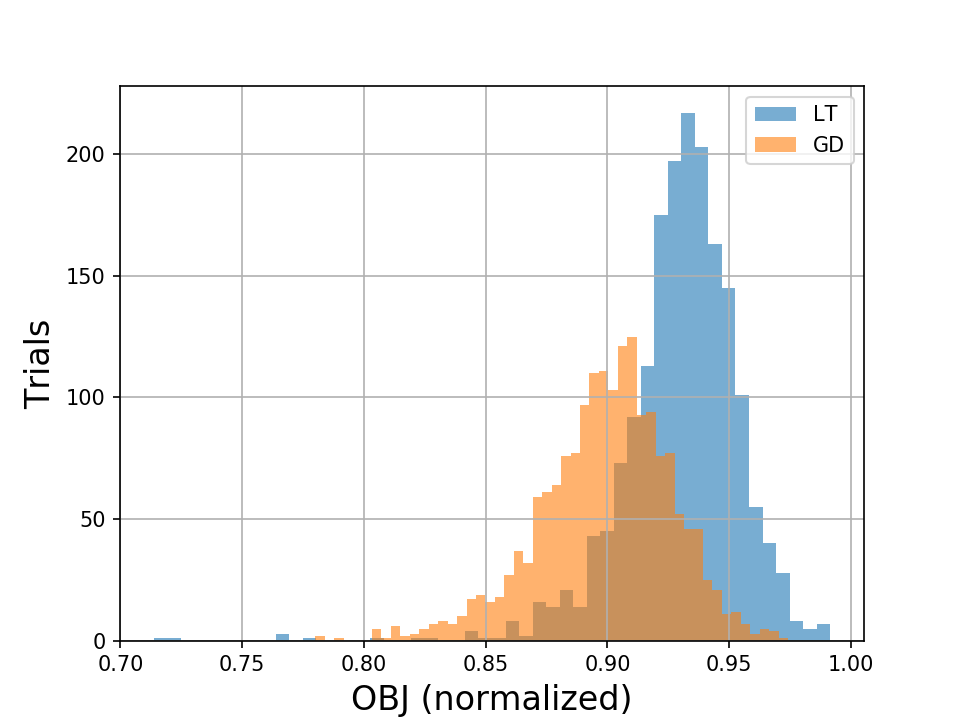}
    \caption{A comparison of \lt{} and gradient descent for a selection of instances. 
    We see that \lt{} produces higher quality optima than gradient descent. 
    For each instance, hyperparameters for both algorithms were tuned to maximize performance.}
    \label{fig:LTvsGD}
\end{figure}

The full algorithm may then be written down:
\begin{enumerate}
    \item Initialize all spins uniformly at random, $v_i\in [-1,1]$.
    \item Apply displacement to spin $v_i\mapsto v_i + c\cdot F_i$ where $F_i = \partial H/\partial v_i$.
    \item $v_i \mapsto \cutoff(\beta v_i)$.
    \item After $p$ rounds, round each spin to its sign, $\pm 1$. 
\end{enumerate}

It is now apparent that GD mirrors \lt{}, with the difference lying in the choice of onsite activation function used: \lt{} uses the tanh function while GD uses a hard cutoff function. 
Both algorithms have identical free parameters $p,c,\beta$ that play the same or similar functional roles in each case. Then, we can compare the performances of these algorithms on the same instances.
In~\cref{fig:LTvsGD}, we see that \lt{} beats GD for the instances shown (and, in fact, all the benchmarking instances studied). This suggests that the specific form of \lt{} that uses a tanh function offers an advantage over a hard cutoff function. In the next section, we motivate this choice further by drawing a parallel between \lt{} and imaginary time-evolution the spins under the problem Hamiltonian.

\section{LT as a discretized, imaginary-time Schr\"odinger evolution}
\label{sec:ImaginaryTime}
Given any initial state $\ket{\psi}$ and Hamiltonian $H$, time-evolution of $\ket{\psi}$ under $H$ is given by the Schr\"{o}dinger equation $d\ket\psi/dt = -iH\ket\psi$. The evolution applies a phase to the eigenstates of $H$ proportional to the energy of the state times time, so that low-energy states rotate slowly while highly excited states rotate fast. An analytical tool often employed to access the low-energy spectrum of $H$ is that of analytic continuation to imaginary time. In this, one replaces the time by an imaginary time parameter $\tau:=it$, and the (unnormalized) imaginary time Schr\"{o}dinger equation reads
\begin{equation}
    \dot{\ket\psi} \equiv d\ket\psi/d\tau = -H\ket\psi\, .
\end{equation}
The formal solution to this equation is $\ket{\psi(\tau)} = e^{-H\tau}\ket{\psi(0)}$. Note that $\ket{\psi(\tau)}$ is unnormalized, but we keep track of the normalization $\mc{N}(\ket{\psi(\tau)}) \equiv \mc{N}(\tau):= \sqrt{\braket{\psi(\tau)}{\psi(\tau)}}$. In the limit $\tau\rightarrow\infty$, and assuming that the ground state of $H$ is non-degenerate, the exponential $e^{-\tau H}$ suppresses contributions from all but the lowest-energy state $\ket{\psi_0}$ of $H$, which implies that $\lim_{\tau\rightarrow\infty} \ket{\psi(\tau)} = \ket{\psi_0}$. 

The normalization $\mc{N}(\tau)$ has $\tau$-dependence 
\begin{align}
    \dot{\mc{N}} &= \frac{1}{2\sqrt{\braket{\psi}{\psi}}}\cdot\paren{\braket{\dot\psi}{\psi} + \braket{\psi}{\dot\psi}}\\
    &= -\frac{\lrang H}{\mc{N}}
\end{align}
where $\lrang H := \bra{\psi}H\ket{\psi}$ is the \emph{unnormalized} expectation value of operator $H$. The normalized expectation value is given by $\lrang{\lrang{H}} := \lrang{H}/\mc{N}^2$.

Next, let $H$ be a Hamiltonian acting on $n$ qubits that is diagonal in the $Z$ basis. Any state $\ket{\psi}$ in this Hilbert space can be mapped to a vector of normalized expectation values of the Pauli operators $Z_i$, where the index $i$ runs over all spins: 
\begin{align*}
    \ket{\psi(\tau)} &\mapsto \paren{\lrang{\lrang{Z_1}},\lrang{\lrang{Z_2}},\ldots, \lrang{\lrang{Z_n}}} \\ &=: \paren{v_1,v_2,\ldots, v_n}\, ,
\end{align*}
where $v_i\in[-1,1]$ is the classical spin variable that tracks the normalized expectation of $Z_i$. The imaginary time-evolution of the spins is given by
\begin{align}
    \dot{v_i} &= \frac{d}{d\tau}\paren{\frac{\lrang{Z_1}}{\mc{N}^2}}\\ &= \frac{-2\dot{\mc{N}}}{\mc{N}^3}\lrang{Z_i} + \frac{1}{\mc{N}^2}\frac{d}{d\tau}\bra{\psi}Z_i\ket{\psi}\, ,\\
    &= 2v_i\lrang{\lrang{H}} - \lrang{\lrang{HZ_i+Z_iH}}\, .
\end{align}
This is essentially an imaginary-time analogue of the Ehrenfest theorem. Since Pauli operators $Z_i$ square to the identity, a diagonal Hamiltonian $H$ can always be written as 
\begin{equation}
    H = R_i Z_i + S_i\, ,
\end{equation}
for every site $i$, for some operators $R_i,S_i$ that are not supported on site $i$. Then, $H Z_i = Z_i H = R_i + S_i Z_i$. Next, we make a mean-field assumption, $\lrang{\lrang{H_{\bar{i}} Z_i}} \approx \lrang{\lrang{H_{\bar{i}}}}\cdot\lrang{\lrang{Z_i}}$, where $H_{\bar{i}}$ is any local operator not supported on site $i$. Then, it follows that $\lrang{\lrang{H Z_i}} \approx \lrang{\lrang{R_i}} + v_i\cdot \lrang{\lrang{S_i}}$, and $\lrang{\lrang{H}} \approx v_i\cdot\lrang{\lrang{R_i}} + \lrang{\lrang{S_i}}$, which gives
\begin{align}
    \dot{v_i} &= 2\paren{v_i^2\cdot\lrang{\lrang{R_i}} + v_i\lrang{\lrang{S_i}} - \lrang{\lrang{R_i}} + v_i\cdot \lrang{\lrang{S_i}}} \\ 
    &= -2\paren{1-v_i^2}\cdot\lrang{\lrang{R_i}}\, .
\end{align}
Next, we make a substitution $u_i := \tanh^{-1} v_i$ which maps the real line into the open interval $(-1,1)$. Then, $\dot{v}_i = 1-\text{sech}^2(u_i)\dot{u}_i = (1-v_i^2)\dot{u}_i$, therefore we can write
\begin{equation}
    \dot{u}_i = - 2\lrang{\lrang{R_i}}.
\end{equation}
Note now that the term $\lrang{\lrang{R_i}}$ is precisely the expected value of the force on spin $i$, since $dH/dZ_i = R_i$. Finally, an imaginary time evolution discretized into small time steps $\delta \tau$ obeys (in mean field)
\begin{equation}
    v_i(\tau+\delta \tau) = - \tanh\paren{2\delta \tau F_i + \tanh^{-1}v_i}.
\end{equation}
This equation bears similarity to the update rule for \lt{}. In fact, for $v_i$ sufficiently small and close to steady state $\bar{v}_i$, we can expand the inverse tangent as $\tanh^{-1}v_i \approx v_i + v_i^3/3 + \ldots \approx - 2\bar{v}_i^3/3 + v_i\cdot(1 + \bar{v}_i^2)$, which looks linear with a modified slope. This reveals a surprising connection between \lt{} and a discretized, mean-field imaginary time evolution of classical spin expectation values.  

Our analysis suggests a generalization of \lt{} to situations where the Hamiltonian is not diagonal in the $Z$ basis. In this context, we can represent each spin $i$ as a 3D rotor $\vec{r}_i = (x_i,y_i,z_i) = \paren{\lrang{\lrang{X_i}},\lrang{\lrang{Y_i}}, \lrang{\lrang{Z_i}}}$ of Pauli expectation values.  Since the Paulis square to the identity, a general spin Hamiltonian $H$ can always be written as
\begin{equation}
    H = P_i X_i + Q_i Y_i + R_i Z_i + S_i\, ,
\end{equation}
for every site $i$, where $P_i, Q_i, R_i, S_i$ are some Hermitian operators that do not take support on site $i$. Then, $H_i Z_i + Z_i H_i = 2R_i + 2S_i Z_i$ (and analogously for $X_i,Y_i$), and therefore 
\begin{equation}
    \dot{x}_i = -2(1-x_i^2)\cdot\lrang{\lrang{P_i}}\, ,
\end{equation}
and similarly for the other coordinates. More succinctly, if we define $\vec{\rho}_i := \paren{\tanh^{-1}x_i,\tanh^{-1}y_i, \tanh^{-1}z_i}$, then the imaginary time evolution becomes
\begin{equation}
    \dot{\vec{\rho}}_i = -2\vec{F}_i
\end{equation}
where $\vec{F}_i = \paren{\lrang{\lrang{\frac{dH}{dX_i}}}, \lrang{\lrang{\frac{dH}{dY_i}}}, \lrang{\lrang{\frac{dH}{dZ_i}}}} = \paren{\lrang{\lrang{P_i}}, \lrang{\lrang{Q_i}}, \lrang{\lrang{R_i}}}$. Then, we can imagine a generalization of \lt{} that discretizes the above equation and simulates the evolution of a 3D rotor. By "rounding" the expectation values of the final state, we arrive at a product state estimate of the ground state. The study of this generalized algorithm will be left as a subject of future work.

\section{Discussion}
The benchmarking of our implementation of \lt{} on the \maxc{} instances gives evidence that \lt{} can perform well in certain practical problem settings. 
We find that the \lt{} hyperparameters can be set using simple rules that obviate the need for a full, global hyperoptimization, making the algorithm particularly lightweight. 

It remains to be seen how well \lt{} fares on problems other than \maxc{}. 
We expect \lt{} to show similar performance in closely related quadratic unconstrained binary (QUBO) problems. 
More generally, we remark that the algorithm itself is specified by a domain relaxation, and a notion of derivative of the objective function with respect to each variable. 
These are minimal requirements found in many optimization problems, for example mixed integer linear programs. 
An interesting open question is whether \lt{} can be adapted for use in these settings as well. The analysis in~\cref{sec:ImaginaryTime} suggests an alternative description of the algorithm as a discretized simulation of imaginary-time dynamics in a spin system. It is interesting whether this picture can be pursued to design improvements or variations to the algorithm, or generalize it to other settings, for instance, on problems like quantum SAT where the problem Hamiltonian is not diagonalizable in any local basis. 

\bibliographystyle{apsrev4-1}
\bibliography{refs}
\end{document}